\documentstyle[12pt,epsfig]{article}
\begin{document}
\begin{titlepage}
\begin{center}
\vspace*{2cm}

{ \Large  \bf  ON THE ELECTROPRODUCTION ON NUCLEI } \vspace{2cm}

\begin{author}
\Large K. Fia{\l}kowski\footnote{e-mail address:
fialkowski@th.if.uj.edu.pl}, R. Wit\footnote{e-mail address:
wit@th.if.uj.edu.pl}

\end{author}

\vspace{1cm}

{\sl M. Smoluchowski Institute of Physics\\ Jagellonian University \\

30-059 Krak{\'o}w, ul. Reymonta 4, Poland}

\vspace{2cm}

\begin{abstract}
Recent data on the electroproduction of hadrons on nuclei are discussed. The effects of the nuclear
absorption are investigated using the Lund model for electroproduction on nucleons. A simple geometrical
model with a minimal number of assumptions and free parameters is shown to describe the data
reasonably well.

\end{abstract}

\end{center}
\vspace{1cm}



\vspace{0.5cm}

\end{titlepage}

\section{Introduction}
   ~~~~Electroproduction of hadrons on the nuclei is considered since long as a favourite
   process to investigate the nuclear absorption effects. On one side, the initial production process may be safely
   assumed to occur on a single nucleon and to be the same as in the electroproduction on a nucleon (proton or neutron)
   in the deuteron, since any screening corrections are negligible for deeply virtual photons exchanged
   in the process. On the other hand, there is a marked difference in the spectra of hadrons from
   these two processes. The obvious effect leading to such a difference is the secondary interaction of the produced
   hadron on its way out from the nucleus \cite{B}.
\par
Recently, interesting data from the HERMES collaboration were published, enlarging our knowledge of the
electroproduction on $N$, $Kr$ and $Xe$ nuclei both for the single spectra \cite{H1} and for the two hadron systems
\cite{H2}.
The direct comparison of data on hydrogen and on heavier nuclei allows to investigate the ratios of spectra for which
many systematic uncertainties cancel.
\par
We present a comparison of these data with a simple model based on
the PYTHIA code \cite{PYTHIA} for the electroproduction on
nucleons and the geometrical scheme for calculating the absorption
effects. Obviously, we are not the first to consider a model of
this type. The unique feature of the Lund description of the
production processes is the definite space-time picture associated
with the generation of hadrons \cite{TS}. Therefore it was
extensively used to describe the absorption effects in the
hadroproduction on nuclei \cite{TC, BC, BG, TC2, BCZ, CZ}. Also
recently various versions of the Lund models were exploited to
analyse the HERMES data \cite{AMP, WANG, FCGM, GF}. However in
these papers the geometrical picture of production from the one
dimensional Lund model was carried directly into the $3D$ case,
and/or coupled to the complicated description of interactions with
"the remaining part" of the nucleus (e.g. the system of transport
equations). It is thus difficult to decide which piece of the
model is responsible for the discrepancies with data.
\par
We discuss a very simple picture, in which only the obvious part of the Lund space-time development is used,
and we supplement it with (equally obvious) pure absorptive effects. Clearly, such a model has a limited applicability.
However, we feel that the determination of the limits of applicability for a maximally simplified model is useful
for any future extensions and more sophisticated descriptions. Moreover, it is instructive to check which features of
the data can be described satisfactorily already in such a simple model and thus, which predictions seem to not depend
on any details of the description. We restrict ourselves to the use of hadronic (and not partonic) degrees of freedom,
since we discuss the low energy data for which the typical $Q^2$ values are small.
\par
In the next section we present our model in detail. Then we discuss the results for the single particle spectra and the
comparison with recent data on hadron pairs. We conclude with a short summary.

\section{The model}

~~~~We are using the Monte Carlo generator PYTHIA 6.203 \cite{PYTHIA} and generate more than a quarter milion of events
per each nucleus, with a correct proportion of $ep$ or $en$ events (the experimental data are shown as ratios of the
distributions from heavier nuclei and from the deuteron).
All the kinematical cuts from two HERMES sets of data are applied, either by setting the proper values of PYTHIA parameters,
or explicitly in the program for the event analysis.
\par
We supplement the ordinary information provided by PYTHIA for each event by the values of one extra parameter from
the generating algorithm. It is the GAM(3) parameter, set for $each$ string break in the PYSTRF procedure, denoting the
proper time $\tau_0$ (time measured in the string rest frame) between the string formation and its break. This time,
corrected for the Lorentz dilatation, is used to calculate the distance between the string formation and string breaking
point in the nucleus rest frame
$$s^0_{form}=\tau_0v_{str}\gamma_{str}.$$
 Then it is attributed as the lower limit for the formation length to the hadron originating
from this break. We allot this length for the charged "stable" hadrons, i.e. pions, kaons and (anti)protons born directly
from the string breaks as well as for the decay products ("children" and "grandchildren") of the hadrons born
directly. For the "last" hadron (or the pair of hadrons) of the event, which combines the "leftover" partons born in the
former breaks, the last calculated proper time of the break is used.\footnote{Thus typically each value of the formation length
is attributed to more than one charged hadron.} We checked that this prescription attributes the
formation length to almost all the charged hadrons produced.
\par
The assumption that the string break point is the "birthplace" of a hadron is known to lead to an overestimation of the
absorption effects. Thus the usual geometrical considerations of the string breaking and hadron formation in the Lund
model \cite{TS} suggest that the formation length should include the distance between the string break and the
"formation point", the first intersection of the lines of partons "born" in the neighbouring breaks \cite{GF}. Therefore
one distinguishes often between two formation lengths: the "production length" and the "yo-yo length" \cite{BG}. The later
choice was found repeatedly to yield too small absorption effects, unless significant absorption cross section for
$partons$ is assumed. Moreover, such an intersection occurs in fact only in the $1+1$ dimensional Lund model; in $(3+1)D$
the lines usually do not meet.
\par
 Thus we keep the idea of a "two stage" formation of hadrons, but consider other ways of accounting for the second stage
 (rearrangement of partons to form a hadron).
We tried two simple choices. In the first case a constant "recombination proper time" $\tau_r$ (of the order of
$0.1\div1$ fm/c in the string rest frame)
is assumed, leading to the extra piece of formation length proportional to the $string$ velocity and the Lorentz factor

$$s_{form}=(\tau_0+\tau_r)v_{str}\gamma_{str}.$$

In the second case for each hadron a "hadronization proper time" $\tau_h$ is assumed, and subsequently dilatated by a
$hadron$ Lorentz factor $\gamma_h$

$$s_{form}=\tau_0v_{str}\gamma_{str}+\tau_hv_h\gamma_h.$$

We found that the last choice overestimates significantly the formation length, leading to the underestimation
of the absorption effects for fast hadrons (the ratios of spectra do not decrease for high hadron momenta).
 Thus this case is not considered in the following. Note that in both cases
we assume no absorption until the hadron is formed (the absorption cross section is zero for the state existing
between the string break and the hadron formation).
\par
For the string creation point we generate the values of the distance from the centre of nucleus $r_0$ and the
azimuthal angle $\theta$ assuming a constant density in volume, i.e. choosing randomly from the flat distributions of
$r_0^3$ between $0$ and $R^3$, and $cos\theta$
between $-1$ and $1$. Here $R$ is the nuclear radius, calculated for each $A$ from the standard formula (in fm units)

$$R(A)=1.12A^{1/3}-0.86A^{-1/3}.$$

To estimate the absorption effects we calculated first the geometrical path from the creation point to the nuclear surface,
 i.e.

$$s_{geom}=r_0cos\theta+\sqrt{R^2-r_0^2sin^2\theta}.$$

            \begin{figure}[h]
\centerline{\epsfig{figure=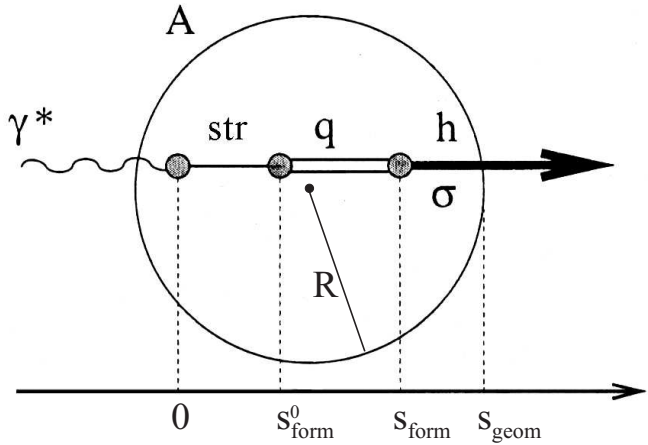, height =6cm}}
\caption{\label{Fig1} {\small \sl The schematic picture of the string formation, string breaking and hadron formation
inside a nucleus. The distances are measured from the string formation point.}}
\end{figure}

Then we evaluate the absorption length $L$ for hadrons inside nucleus
from the absorption cross section $\sigma$ and nuclear density $\rho$, $L=(\sigma \rho)^{-1}$. The inelastic cross sections
in the energy range corresponding to the HERMES conditions are about $50$ mb for antiprotons, $30$ mb for protons,
$20$ mb for pions and negative kaons and  about $12$ mb for positive kaons. This yields the values of absorption length
of $1.2 \div 5$ fm. With these values we may calculate the survival probability

\begin{displaymath}
P=\left\{ \begin{array}{ll} exp(-s'_{geom}/L) & \textrm{if$~s'_{geom}=s_{geom}-s_{form}>0$},\\
1 & \textrm{if$~s'_{geom}<0$}, \end{array} \right.
\end{displaymath}
\noindent
which is used as the weight to produce the spectra from
nuclei. The ratio of spectra with weights and without weights (i.e. with weight equal one) is to be compared
with the experimental ratios of spectra from nuclei and the properly normalized spectra from nucleons.
The geometrical picture showing the quantities defined above is presented in Fig.1 for the case of a hadron formed inside
the nucleus.
\par
In these calculations the nuclei were modelled by a very simple description, the so-called
"hard sphere", in which the nuclear density is constant within a sphere of radius $R$, and zero outside.
In a more realistic model the nuclear density depends on the radial coordinate e.g. by the standard
("Saxon-Woods") formula

$$\rho(r)=\frac{\rho_0}{1+exp[(r-R)/\Delta]},$$
\noindent
where $\rho_0=0.17~fm^{-3}$ and $\Delta=0.54~fm.$ To check if the geometrical approximation is reliable
we have repeated the calculations with $s'_{geom}$ replaced by $s'_{eff}$

$$s'_{eff}=\int_d^{\infty}ds\biggl[1+exp\frac{\sqrt{s^2+r_0^2sin^2\theta}-R}{\Delta}\biggr]^{-1}$$
where $d=-r_0 cos \theta +s_{form}$. Another modification was the
selection of the string creation point not just from the inside of
the sphere, but with $r_0$ weighted with the Saxon-Woods
distribution.

The ratios of $z$-distributions resulting from the weights calculated
in the discussed cases for all nuclei are hardly distinguishable. Since the generation of $r_0$ values from a
non-uniform distribution is more time consuming, we use in the following the model with variable nuclear density, but
with the simple generation of the $r_0$ values.

\section{Absorption for the single hadron spectra}

~~~~~~The HERMES collaboration has published precise data for single charged hadron electroproduction on the deuteron,
nitrogen and krypton gas targets for the positron energy of $27.6$ GeV \cite{H1}. The relative energy in the lab frame $z=E_h/\nu$
was used to define the spectra in the range $0.02<z<1.0$. The ratio of the properly normalized spectra

$$ R^h(z,\nu,p_t^2,Q^2)=\Big(\frac{N_h(z,\nu,p_t^2,Q^2)}{N_e(\nu,Q^2)}\Big)_A\Big/
\Big(\frac{N_h(z,\nu,p_t^2,Q^2)}{N_e(\nu,Q^2)}\Big)_d$$

\noindent
reflects the nuclear effects for the
light (nitrogen, $A=14$) and medium-size (krypton, $A=84$) nuclei. The kinematical cuts for the momentum transfer
squared, $Q^2>1.0$ GeV/$c^2$, hadronic invariant mass $W>2$ GeV and the positron relative energy loss  $y=\nu/E<0.85$
were used.
\par
These results were compared in the HERMES paper \cite{H1} with two absorption models \cite{AMP,WANG}. One of
them \cite{AMP} visibly overestimates the absorption effects for nitrogen, and the other one \cite{WANG}
predicts a too steep $z$-dependence for the krypton data. Obviously, none of the models describes the
excess of hadrons at small $z$, which is probably due to the production in secondary collisions.
\par
We compare the same data with the predictions of our model specified in the previous section. For the
formation length defined by the breaking point of the string, $s_{form}=\tau_0\gamma_{str}v_{str}$, the
absorption effects are too strong, especially for the krypton data. Thus it seems necessary to introduce the
"recombination proper time", as described above.

            \begin{figure}[h]
\centerline{\epsfig{figure=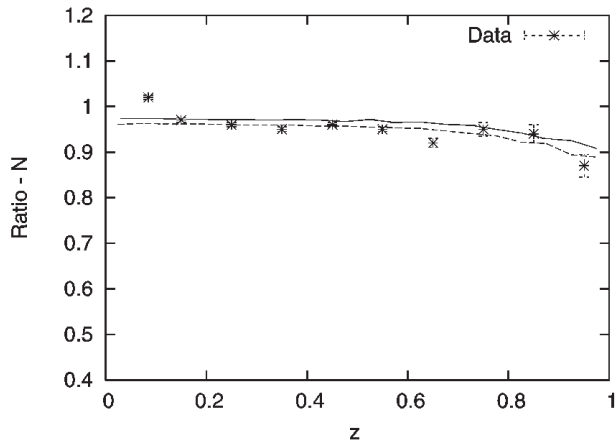, height =6cm}}
\caption{\label{Fig2} {\small \sl The experimental ratio of the
charged hadrons $z$-spectra for nitrogen and deuterium \cite{H1}
(stars) compared with the model calculations for $\tau_r=0.7$ fm/c
(broken line) and for $\tau_r=0.8$ fm/c (solid line).}}
\end{figure}

            \begin{figure}[h]
\centerline{\epsfig{figure=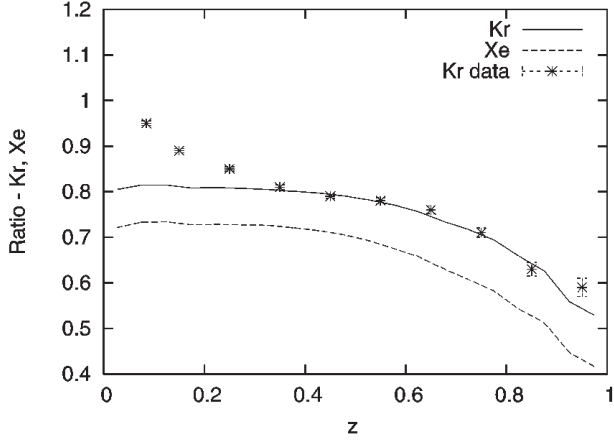, height =6cm}}
\caption{\label{Fig3} {\small \sl The experimental ratio of the
charged hadrons $z$-spectra for krypton and deuterium \cite{H1}
(stars) compared with the model calculations for $\tau_r=0.8$ fm/c
(solid line). The model calculations are shown also for xenon
(broken line).}}
\end{figure}

The value of $\tau_r$ was not fitted to any particular set of data, but we found that the value of $\tau_r=0.7 \div 0.8 $
fm/c gives a good description of the data for nitrogen for $z>0.1$ (Fig.2).  The dependence of the results on this value is
not very strong.  In the following we use only $\tau_r=0.8$ fm, which describes better the data for krypton for $z>0.3$
(Fig.3).

\par The good agreement with data of a simple model with only one free parameter $\tau_r$ for $z>0.3$ is remarkable.  One
should note that both the $z$-dependence and $A$-dependence of data seems to be correctly described using the geometrical
picture provided by the Lund model.  Obviously, it is desirable to get similar data for more than two values of $A$, e.g.
for xenon, for which the predictions are also shown in Fig.3.
\par It is
obvious that we cannot expect an agreement of the model calculations with the data for small values of $z$.  We have
considered merely the absorption effects (and get in fact the results quite similar to other authors cited in
the HERMES paper \cite{H1}, although our model gives a better description of the data for large values of $z$).
We expect that the secondary collisions of the hadrons originating from the string
breaking with other nucleons produce extra hadrons.  This enhances the spectra for low $z$ values and may even result in
the values of $R_h$ above one, as seen in the first bin of the nitrogen data.  Such an effect may be directly
incorporated
into the model \cite{CS}, or combined with the absorption effects in the transport equations \cite{FCGM,GF}.
However, such refinements are rather arbitrary and involve many extra parameters.  Therefore we feel that it is
useful to establish the range of $z$ for which a simple absorption model does not require any such modifications. It seems
to be almost the full $z$ range for light nuclei, as nitrogen ($z>0.1$), and visibly less for heavier nuclei,
as krypton ($z>0.3$).
It would be interesting to check if the same range applies as well for other nuclei, e.g. xenon.
\par
Till now we have discussed the data for all charged hadrons. The HERMES experiment is able to identify particles and the
corresponding data were also published \cite{H1}. The data for pions resemble closely those for all charged hadrons, as
pions are most copiously produced. The data for negative kaons do not agree with the predictions of the absorption model
\cite{AMP}. In particular, this model predicts almost negligible difference between the positive and negative kaons,
in clear disagreement with data.
\par
In our model one may expect different absorption effects for pions and kaons for two reasons. First, the absorption
cross sections are different, as discused in the previous section. Second, the different masses may lead to different
values of the only free parameter,
the recombination time $\tau_r$. We found that a reasonable agreement with data for kaons at $z>0.35$ is achieved for
$\tau_r=0.4$ fm/c, half of the $\tau_r$ value for pions. Let us note that by the uncertainty relation these values
correspond to the energies of about $0.3$ GeV for pions and $0.6$ GeV for kaons, close to their typical transverse masses.
The data are shown together with our predictions in Fig.4.
            \begin{figure}[h]
\centerline{\epsfig{figure=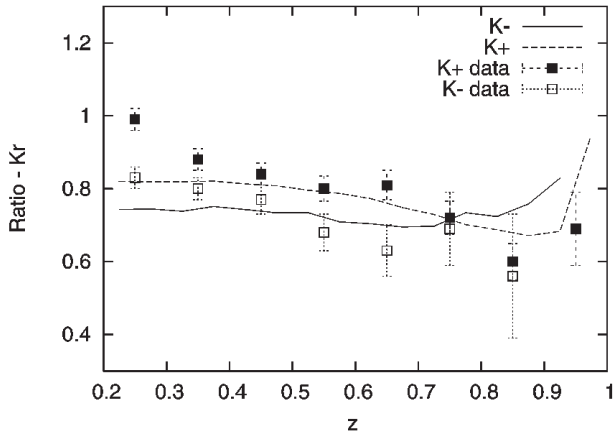, height =6cm}}
\caption{\label{fig4} {\small \sl The ratio of normalized yields of $K^+$-s (solid squares) and $K^-$-s (open squares)
for Kr and D nuclei \cite{H1}. Our model predictions are shown as broken and solid lines, respectively.}}
\end{figure}

\section{Two hadron spectra}

~~~~~Recently, the HERMES collaboration has extended their previous data by publishing precise results on the conditional
$z$-spectra for charged hadron electroproduction on the deuteron, nitrogen, krypton and xenon gas targets for the same
positron energy of $27.6$ GeV \cite{H2}. Kinematical cuts were the same as for the single spectra \cite{H1}.
The events were selected by the condition that one of the charged hadrons has $z_1>0.5$.
The ratio of the properly normalized spectra for other hadrons
$$R_{2h}(z_2)=\Big(\frac{dN^{z_1>0.5}(z_2)/dz_2}{N^{z_1>0.5}}\Big)_A\Big/\Big(\frac{dN^{z_1>0.5}(z_2)/dz_2}
{N^{z_1>0.5}}\Big)_d$$
(of the same charge sign as the selected one) were shown for $0.05<z_2<0.5$.
\par
 The results were compared with the predictions of absorption models \cite{AMP,WANG} and with the
predictions of a model including the production processes by means of the BUU transport equations \cite{FCGM}. The absorption
models overestimate the absorption effects, especially for heavier nuclei, in all the considered $z$-range.
The transport model fares better, but it does not reflect two most striking features of the data: the similarity of the
results for all the nuclei and the apparent disappearance of the absorption effects for $z_2 \rightarrow 0.5$.
            \begin{figure}[h]
\centerline{\epsfig{figure=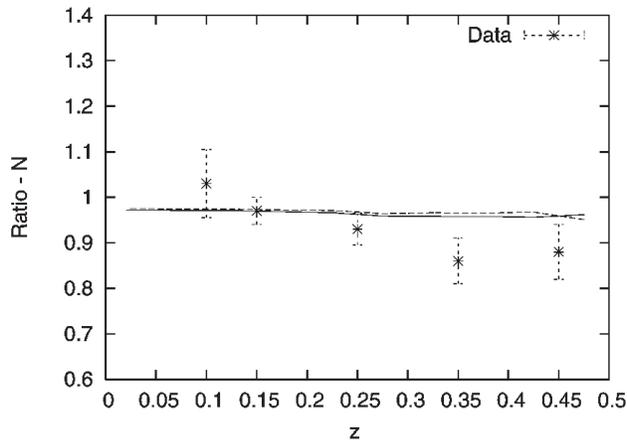, height =6cm}}
\caption{\label{fig5} {\small \sl The experimental ratio of the
conditional charged hadrons $z$-spectra for nitrogen and deuterium
\cite{H2} (stars with errorbars) compared with the model
calculations for $\tau_r=0.8$ fm/c (broken line) and the results
of the event generation without proper correlations (solid
line).}}
\end{figure}

\par We compare the data with the same absorption model which we used to describe the single spectra (with the value of
$t_r=0.8$ fm).  Let us note that a simple selection of PYTHIA events for $ep$ and $en$ collisions with the value of
$z_1>0.5$ for one of the hadrons is not sufficient to model the nuclear collisions.  Due to the strong nuclear absorption
of the fast "trigger hadron" the cut on $z$ discriminates against the selection of a hit nucleon in the "front part" of the
nucleus.  Therefore, on average, the other hadrons will be absorbed less strongly than in a typical event.

\par
 In the Lund models this effect is obviously present.
The observation of a fast hadron in electroproduction on heavy nucleus enhances the probability that the
absorption of $all$ the hadrons in this event is below the average. This happens because the absorption depends both on
$s_{form}$ (which may be different for each hadron) and $s_{geom}$, which is common for all the hadrons from a given event.
Therefore a proper procedure to generate events for the conditional $z$-distributions is to calculate for each event with
a fast hadron ($z_1>0.5$) the probability $P_{na}$ that this hadron is not absorbed before leaving the nucleus. Then one
generates a random number $r$ from the range $0\div 1$ and accepts the event only if $r<P_{na}$.

            \begin{figure}[h]
\centerline{\epsfig{figure=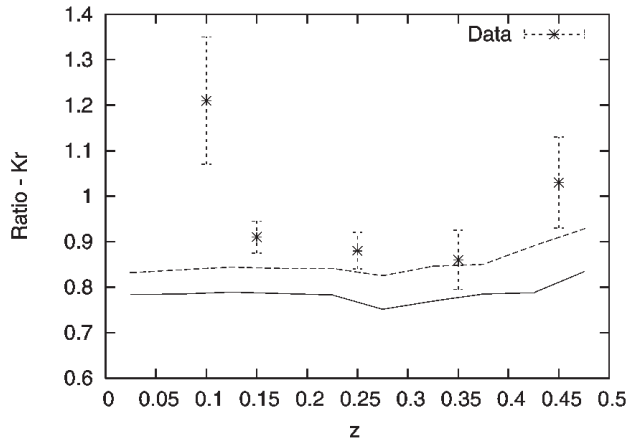, height =6cm}}
\caption{\label{fig6} {\small \sl The experimental ratio of the
conditional charged hadrons $z$-spectra for krypton and deuterium
\cite{H2} (stars with errorbars) compared with the model
calculations for $\tau_r=0.8$ fm/c (broken line) and the results
of the event generation without proper correlations (solid
line).}}
\end{figure}
            \begin{figure}[h]
\centerline{\epsfig{figure=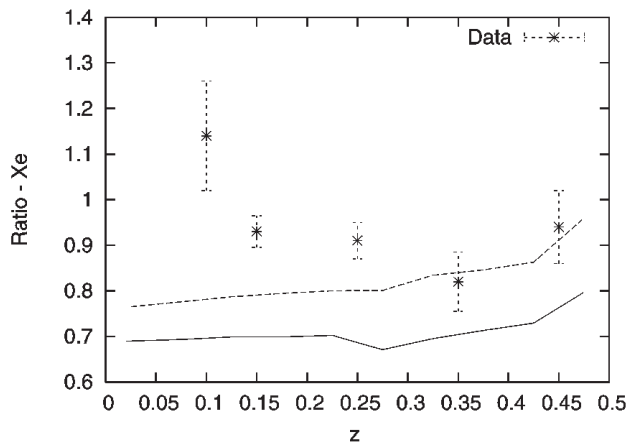, height =6cm}}
\caption{\label{fig7} {\small \sl The experimental ratio of the
conditional charged hadrons $z$-spectra for xenon and deuterium
\cite{H2} (stars with errorbars) compared with the model
calculations for $\tau_r=0.8$ fm/c (broken line) and the results
of the event generation without proper correlations (solid
line).}}
\end{figure}

\par In figures 5-7 we show both the predictions of our model with the correlation taken into account, as described above,
 and without this correlation (showing just the absorption for the hadrons other than the fastest one).  We see that the
 nitrogen data show stronger absorption than expected (and stronger than seen in the single spectra, c.f.  Fig.2), but the
 disagreement is in the worst case on the level of two standard deviations.  The data for krypton and xenon are marginally
 consistent with the model for $z_2>0.2$ (Kr) or $z_2>0.3$ (Xe).  The important point is that the model with correlations
 is much closer to the data.
\par Let us remind once more that for lower $z$ the production processes (neglected in our model) are probably important.
The results for the model neglecting correlations overestimate significantly the absorption effects for heavier nuclei.
They are quite similar to the results of the absorption models shown together with the HERMES data \cite{H2}.

\section{Conclusions}

~~~~~We have investigated
 the electroproduction of hadrons inside the nuclei using the PYTHIA
 generator.
The results from the recent HERMES experiment \cite {H1,H2}
 are compared with the calculations done using a very simple scheme for the absorption effects. The ratios of
 single particle spectra for all charged hadrons are quite well described for the values of $z$ above $0.1$ (N)
 or $0.3$ (Kr). For the lower $z$ values our model predicts a flat $z$-dependence (as any absorption model),
 in disagreement with data.
 The flavour dependence of the absorption effects in single spectra is resonably well described assuming that the
 recombination time is proportional to the inverse of the hadron transverse mass. The ratios of the conditional spectra
 for the events containing one hadron with large
 $z$ are also consistent with data for the values of $z$ above $0.1$ (N) or $0.3$ (Kr, Xe).  In particular, the
 similarity of the conditional spectra for all nuclei and the increase of ratios for maximal possible $z_2$
 seems to be well understood in the absorption model with the PYTHIA generation mechanism. These effects result from the
 correlation induced by the bias in the position of a string creation for the events with a very fast hadron.
 \par
 Our results
 suggest that the space-time production picture in the Lund model coupled with a simple absorption model leads to a
 satisfactory description of the large part of the experimental data.

\section{Acknowledgements}

~~~~~ We are grateful to A. Bia{\l}as for fruitful discussions and in particular for the suggestion to investigate
the flavour effects. We thank A. Kota{\'n}ski for reading the manuscript and for helpful
 remarks. This work was partially supported by the research grant 1 P03B 045 29 (2005-2008).

\end{document}